# Giant absorption of light by molecular vibrations on a chip

A. Karabchevsky,[1,*] and A. V. Kavokin,[2,3‡]

[1]Department of Electrooptic Engineering, Ben-Gurion University, 84105, IL
[2]Department of Physics and Astronomy, University of Southampton, SO17 1BJ, UK
[3]CNR-SPIN, Viale del Politecnico 1, I-00133 Rome, Italy

**Vibrational overtone spectroscopy of molecules is a powerful tool for drawing information on molecular structure and dynamics[1]. It relies on absorption of near infrared radiation (NIR) by molecular vibrations. Here we show the experimental evidence of giant enhancement of the absorption of light in solutions of organic molecules due to the switch from ballistic to diffusive propagation of light[2-6] through a channel silicate glass waveguide. We also experimentally address a dynamics of absorption as a function of time of adsorption of the organic molecules on a waveguide. The observed enhancement in diffusion regime is by a factor of 300 in N-Methylaniline and by factor of 80 in Aniline compared to the expected values in the ballistic propagation of light in a waveguide. Our results underscore the importance of a guide surface modification and the disordered[6] molecular nano-layer in enhancement of absorption by amines on engineered integrated system[7].**

Aromatic amines, such as N-Methylaniline (NMA) for which Aniline is a prototypical molecule, are widely used in biology and medicinal chemistry. In particular, they constitute a crucial component of many drugs, pesticides and explosives. Since most amines are basic, salts can be generated to facilitate water solubility or solubility in other vehicles for drug administration[8]. Historically, the amines were also largely employed in conjugating aromatic rings for applications in the dye industry.

Nowadays, the interest in the vibrational spectra of amines has been boosted by the potential application of molecular vibration spectroscopies in the detection of explosive materials and diagnostics of psychoactive stimulants based on amines[9]. The most wide-spread method of vibration spectroscopy of molecules relies on the operations in a mid-infra-red (Mid-IR) spectral range. Its main disadvantages are the needs for bulky and expensive equipment and for significant quantities of sample solutions in order to extract high precision data on spectral positions of vibrational modes. From the perspective of point of care applications, there is a need for a micro-size molecular spectroscopy set-up suitable for operations with low volume samples. Here we demonstrate a near-infra-red (NIR) spectroscopy laboratory on a chip capable of high precision measurements of resonant NIR absorption spectra of organic molecules[10]. We have performed the transmission spectroscopy measurements on samples containing Aniline and NMA molecules. The absorption efficiency of the solution placed in a close vicinity to a silicate glass channel waveguide has been found to be increased by a factor of up to 300 compared to the theoretical value obtained assuming the ballistic propagation of NIR in the waveguide channel. We interpret this dramatic enhancement in terms of the switching between ballistic and diffusive propagation of light[11] induced by the resonant scattering of light by organic molecules.

Figure 1a shows an overtone vibrational spectrum of the pure NMA and of the NMA / hexane (Hex) mixtures photographed in Figure 1e. The inset (Figure 1b) shows the absorption bands of NMA due to the (N-H)-bond-stretching band around 1.5 μm and the aryl (C-H) overtone band around 1.65 μm in an overtone region $\Delta V = 2$. The skeletal molecular structure of N-Methylaniline is shown in Figure 1c.

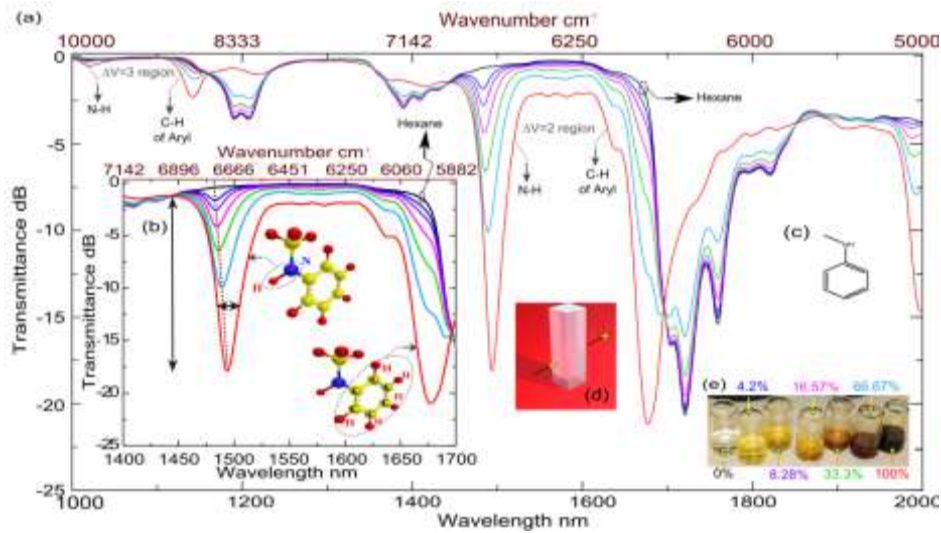

**Figure 1| An overtone vibrational spectrum of the organic molecule N-Methylaniline. a,** Transmittance spectra of pure N-Methylaniline (NMA), pure hexane (Hex, black curves) and NMA diluted in Hex measured by Jasco V570 spectrophotometer recorded at room temperature of 21±2°C and converted to *dB* units. The inset **b** shows the absorption bands of NMA due to the (N-H)-bond stretching band around 1.5 μm and the aryl (C-H) overtone band around 1.65 μm in the ΔV = 2 region. Shift in location of the N-H band with respect to increase in concentration indicated by the dashed line. Arrows are pointing to a change in the signal strength and width. Skeletal molecular shape of NMA is shown in the inset **c**. The quartz cuvette pathlength is 5 mm and the measurement was performed at normal incidence as drawn in **d** with air as reference. **e,** Photograph of the pure NMA and diluted mixtures prepared for the measurement with colored concentration values corresponding to the colors of the spectral curves.

We emphasize that although broadband illumination probes many electronic states, our waveguide, optimised for the single mode regime around 1.5 μm, only supports overtone electronic states in the 1$^{st}$ overtone region (ΔV=2) corresponding to the transition from the ground molecular state to the second size quantized energy state. Note that in the vicinity of the 1$^{rst}$ and second overtones the molecular potential acting upon the electronic excitation is nearly parabolic (see Figure 1f).

Integrated optics or 'thin film guided wave optics' is harnessing from miniaturization due to the manipulation of the optical waves guided in a thin film deposited on a lower refractive index substrate[12]. Light can be guided in a doped substrate if the dopant induces an increase of its refractive index. Here we used diffused K+ ions in the silicate glass to form the guiding layer with a higher index. The optical mode in such structures can be also confined laterally due to the dopant concentration variation in the plane of the structure, see equation (1). Therefore such a structure prevents lateral diffraction of NIR to sides of the guided region. In diffusion based waveguides, lateral confinement is governed by the width of the dopant distribution as in equation (1). Light guiding within a thin film illuminated by the monomode fiber[13] enables transmittance spectroscopy studies of the media surrounding the wave-guide due to the evanescent tails of the propagating light mode. Here we use fibre-compatible ion-exchanged waveguide in silicate glass operating in the monomode regime at the wavelengths of the 1$^{st}$ overtone region of anilines ($\Delta V=2$). Figure 2b shows a photograph of the waveguide with a polydimethylsiloxane (PDMS) liquid reservoir (see Methods section for fabrication details) with input/output fibres guiding green and red laser lights for demonstration of potential multi-array detection. During the experiment, a 0.1 mL droplet of 66.7% NMA in Hex mixture has been dripped onto the liquid reservoir with a micropipette for spectral analysis. Figure 2c shows the NIR transmittance spectra of the waveguide which is in contact with the solution of NMA molecules recorded using the set up shown in Figure 2a after about 30 min from the first contact of the solution with the structure. The signals were recorded every 1 min. An immediate drop in transmittance has been observed using our device after the first contact with the mixture. Figure 2c shows the time evolution of the relevant transmittance spectra.

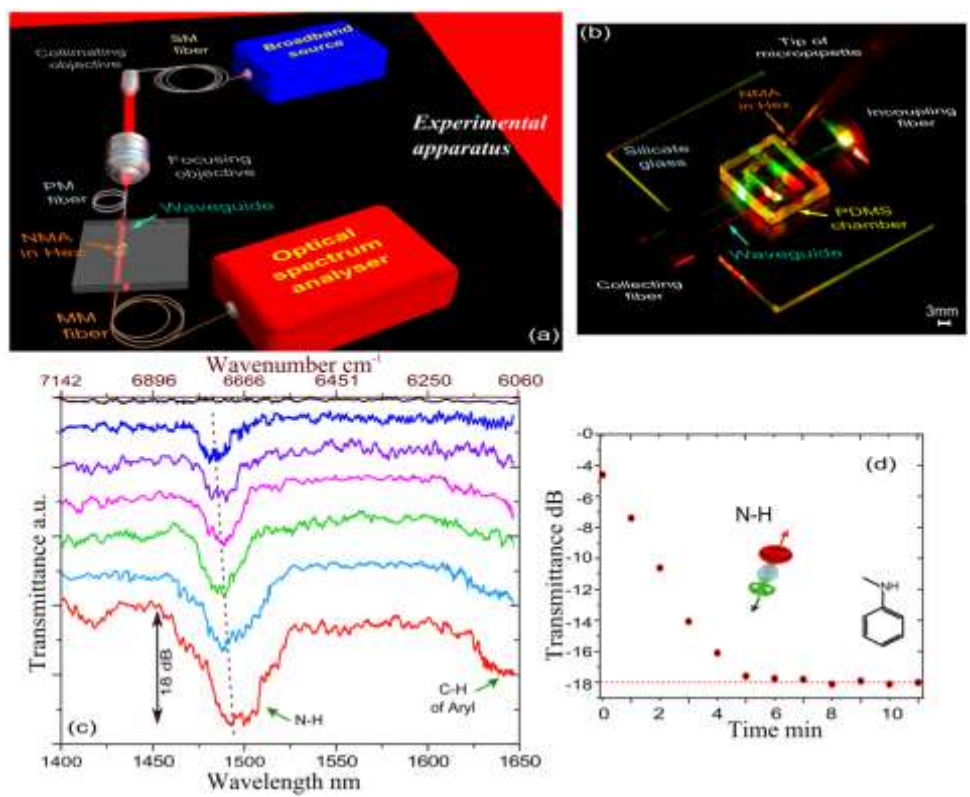

**Figure 2| Experimental set-up and recorded spectra of NMA adsorbed to the waveguide. a**, Artist's impression of the experimental set-up based on butt-coupled broad-band illumination of a waveguide system by a supercontinuum NIR source and recording of the transmittance spectra with use of OSA (See Methods). Tear-sized droplet of a liquid has been dripped onto the waveguide for the spectroscopy analysis. **b**, photograph of the chip during the measurement, NMA has been dripped in Hex solution onto PDMS liquid reservoir using micropipette; in the photograph, two waveguides are illuminated by coherent red and green light sources for demonstration of potential multi-array detection. **c**, Transmittance spectra (solid curves) recorded using the waveguide while dripping volume of 0.1 mL and concentration of 66.7 % NMA diluted in Hex mixture. The spectra are registered at time intervals of about 1 min with time increasing from the top to the bottom of the figure. The blue solid curve corresponds to the min 0 in the subplot (d). The red curve corresponds to the transmittance decrease by 18 *dB*. Other curves have been shifted vertically for clarity. N-H and aryl C-H absorption bands areindicated by the arrows; **d**, Experimental transmittance modulation depth of (N-H)-bond stretching band shown as a function of time.

The effective concentration of NMA in the vicinity of the waveguide surface increases with time and eventually achieves the maximum loss of about -18 *dB* (Figure 2d) which corresponds to the loss measured for the pure NMA sample in a cuvette (Figure 1b). NIR measurements from cuvettes of different NMA/Hex mixtures illustrate that the presence of Hex in the mixture shifts the N-H peak toward shorter wavelengths (Figure 1a-b and Figure 2c). NIR data recorded on a waveguide from the droplet of 67% NMA diluted in Hex shows that in 35 minutes the centre of the N-H peak shifts from 14.75 μm to 14.9 μm, indicating that the concentration of NMA within the penetration depth of the evanescent NIR field increases with time. This signal then remains steady, demonstrating that a compact organic layer has formed near the surface and the Hex has been pushed beyond the evanescent field region (Figure 2d). Complementary surface analytical techniques were used to characterize the formed layer: ellipsometry and contact-angle measurements (see Supplementary Section 3.3).

Let us emphasize the magnitude of the observed effect. Channel waveguides were designed for the monomode operation and high sensitivity at a wavelength of 1.5 μm using the finite element method (FEM) and employing an approximate diffusion profile of potassium ions in glass[14] (Figure 3c) according to equation (1) with the maximum core index of 1.5105.

$$n(x,y) = n_s + \Delta n\, erfc\left(\frac{y}{dy}\right) e^{-x^2/dx^2} \tag{1}$$

With substrate refractive index of $n_s$=1.5013 at 1.496 μm. The coordinate *y* in equation (1) is the distance from the surface and *x* is the lateral distance from the waveguide symmetry axis; Δn is the maximum increase of the refractive index due to the dopant, which is 0.0092 for TM- and 0.008 for TE-polarised modes, *dx* is the profile depth and *dy* is the profile half-width. Figure 3b shows the calculated transmittance of our device, in ballistic regime, in the wavelengths range of N-H stretching band of NMA in region ΔV=2. The resonant scattering leads to switching from the regime of ballistic to diffusive propagation of light[2,3] - a phenomenon which is well known e.g. in most biological tissues, where photon propagation is quickly randomized due to the elastic scattering. This results in a background loss in optical transmittance and a strong delay in the transmission of light around the resonance[4]. It is important to underline that in the diffusion regime photons propagating through a waveguide are absorbed with a higher probability than in the ballistic regime, in general. This tendency has been recently established by the experiments on exciton-polaritons propagating through crystal slabs[11]. Qualitatively, the longer time a photon spends in an absorbing medium the higher chances it has to be absorbed.

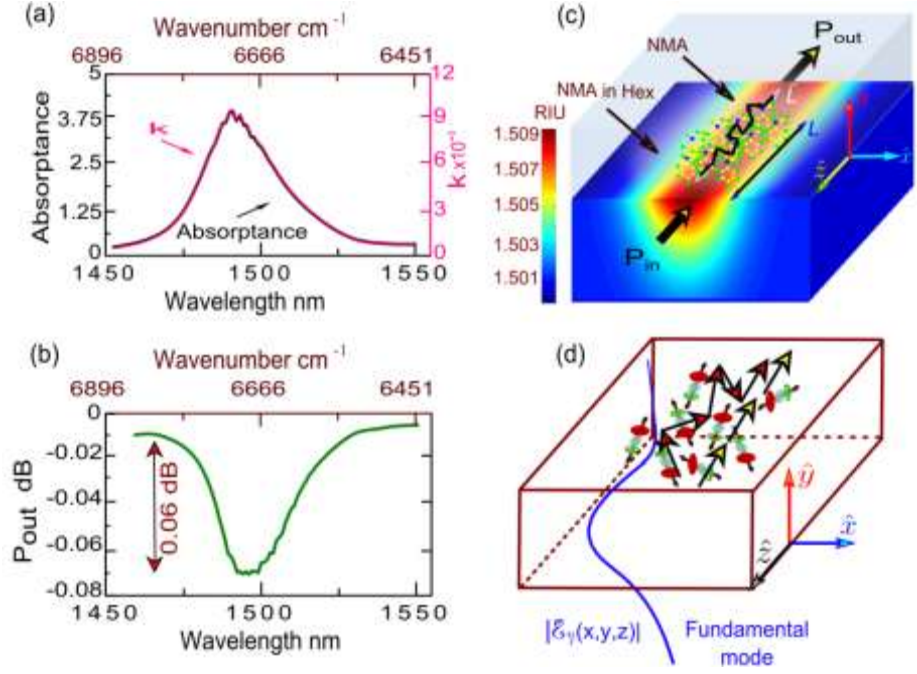

**Figure 3| Transition from ballistic to diffusive 'Slow light' propagation in a guide due to the diffusion by NMA molecules.**
**a,** Measured absorbance of a pure NMA and the corresponding calculated κ; **b,** Calculated transmittance in the ballistic regime with pure NMA; **c,** Calculated refractive index profile n(x,y) for TE polarization of the waveguide using COMSOL Multiphysics 4.3b, photon ballistic pathlength $L$ and photon elastic scattering mean-trajectory $\tilde{L}$ are shown schematically; **d,** Schematic showing the ballistic (yellow arrows) and diffusive (red arrows) pathways of NIR photons in the waveguide (substrate region is not shown for simplicity) with organic molecules adsorbed to the surface. The blue curve describes the electric field intensity spatial distribution of the fundamental mode in a guide $|\mathcal{E}_y(x,y,z)|$.

We have estimated the absorption amplification due to the diffusive propagation of light through the waveguide. The extinction coefficient $\kappa(\lambda)$:

$$\kappa(\lambda) = log_e(10^{A(\lambda)}) \frac{\lambda}{4\pi L} \qquad (2)$$

has been deduced based on the Beer-Lambert low and measured absorption (A) data using a spectrophotometer with 10 mm path-length (Figure 3a). Transmitted power through the guide was calculated as $P \approx$ -0.06 *dB* for ballistic propagation of light through the guide (Figure 4b) with $L \approx$ 3 mm where:

$$P = -10 log(e^{-4\pi L \kappa/\lambda}) \qquad (3)$$

Here $L$ is the length of the contact region of the waveguide with the liquid. The switch to the diffusive propagation, results in the replacement of $L$ by $\tilde{L} \gg L$. In this case, the outgoing signal is reduced by a factor of $\tilde{L}/L = (P'_{out}/P_{out})$, where $P'_{out}$ is a measured power, $\tilde{L}$ is the estimated mean trajectory of a photon in the diffusive medium, and $P'_{out}$ is a transmitted power calculated using *L* as the trajectory length in the ballistic regime. Taking $\kappa=9.4314 \times 10^{-5}$, $\lambda=1.496$ μm, $P'_{out}=-18$ *dB*, $P_{out}=-0.06$ *dB*, we obtain $\tilde{L} \cong 0.9$ m. This differs from the theoretical expectation for the ballistic propagation case by a factor of about 300. We explain this giant discrepancy by two crucial effects: (1) the chemical adsorption of organic molecules to the surface of the guide, their reorganisation and the consequent formation of a dense structured layer; and (2) the physical effect of resonant scattering of NIR by disordered film of organic molecules. In the modelling we performed using the Maxwell equation solver, the increase in the effective thickness of adsorbed molecular layer as a function of time was insufficient to explain the giant modulation of absorption within the ballistic model. We argue that the resonant scattering of light by organic molecules placed in the close vicinity to the waveguide is responsible for the observed effect. The scattering causes the photon propagation direction to change randomly as shown schematically in Figures 3c-d. The Maxwell solver is no more convenient for description of light propagation in this regime. Instead, a diffusion equation for photons should be used (see e.g. equation (4) in Ref. 4 and Supplementary Information Section 1-2).

An increased absorption and delay of light near the resonance was observed both in GaN and in ZnO material[Shubina1, Shubina2]. Here, the diffusive regime is established because of the adsorption of organic molecules on the surface of the waveguide (see Supplementary Information Section 3). The formation of an 8 nm thick organic layer has been detected by the ellipsometry. Attraction of NMA (index of $n_{NMA}=1.571$) to the modified surface of our device is driven by the surface tension between the non-polar solvent, Hex, and the polar waveguide surface. The polar N-H bond in NMA can align with the polar surface, and the non-polar benzene ring aligns with the solvent, relieving the surface tension (Figure 4a). Since benzene rings also tend to stack together, and N-H bonds in different NMA molecules will form hydrogen bonds with each other, a multilayer structure can form much like a lamellar liquid crystal[15]. Contact angle measurements shown in Figure 5b confirm the waveguide surface modification due to adsorption. In addition to the aromatic ring, NMA molecules contain a methyl group such as in Toluene (index of $n_{Toluene}=1.4968$) and an amine group such as in Aniline (index of $n_{Anl}=1.5863$). The spectral responses of those molecules on our device are shown in Figure 4c. However, despite the fact that the N-H overtone absorption band of Aniline in the NIR spectrum has stronger absorption compared to NMA, it has a less pronounced transmittance resonance on a guide. This observation can be explained by the increased hydrogen bonding existing between the amine molecules in aniline, which makes it less energetically favourable for them to reorganise and align with the surface (Figure 4d)[16]. The N-H overtone bands may be shifted, as in Aniline, by changing the solvents and temperature conditions. The effects of hydrogen bonding on the mechanical anharmonicity of the N-H stretching vibration have been reported[17,18]. Weak hydrogen bonding appears to reduce the anharmonicity of the vibration resulting in the broadening of the overtone absorption peak width[17,19].

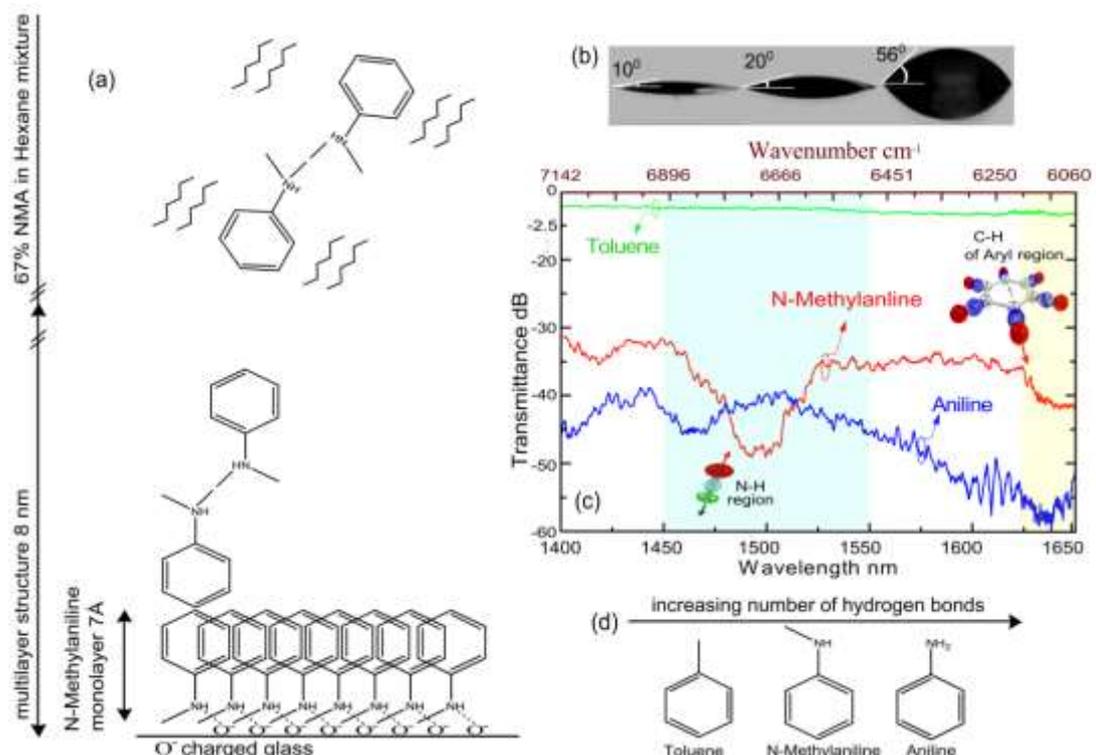

**Figure 4| Responses from benzene molecules with different substituents demonstrating the influence of hydrogen bonding. a,** schematic illustrating a multilayer structure of NMA formed on the surface of the waveguide in the 67%

NMA/Hex mixture. Dashed lines represent hydrogen bonding between two moieties; **b,** surface modification measurement of planar waveguide with the contact angle ∠ of the water droplet on the clean glass as ∠10°, ∠20° close to the immersion area strongly exposed to the solution vapour and a droplet of water on the immersed surface with a contact angle of ∠56°; **c,** NIR transmittance spectra of Toluene, aniline and NMA on our device; **d**, molecules used in this study ordered according to the number of hydrogen bonds they can form with other moieties (Toluene: 0, NMA:1 and Aniline:2)

To conclude, due to the chemical adsorption process, a diffusive layer rich with organic molecules settled off the solution has been formed at the surface of the waveguide. This layer is responsible for the switch from ballistic to diffusive propagation of NIR radiation in the waveguide and the giant absorption detected. Compared to the demonstrated detection of N-Methylaniline on planar integrated optical component[20,21], a ring resonator[22] or microtapered fiber[20] our system achieved dramatically stronger absorption signal of N-H overtone at around 1.5 mm wavelength that harnesses cheap waveguide fabrication procedure optimised for NIR (telecommunication) wavelengths. Detection of the primary amine, Aniline on a waveguide was shown for the first time. The affinity of aromatic amines to the surface of an integrated optical device makes our approach particularly suitable for NIR detection and spectroscopy studies of amine based group materials such as anilines important for pharmaceutical and other chemical industries[23].

**Methods**

**Sample fabrication and characterization**.
The waveguides were fabricated by photolithographical patterning of 250 nm thick hard mask on silicate glass. This results in the stripe openings of 6 μm width. Then, the glass was immersed in a $KNO_3$ molten for 11 hr at 395°C. After ion-exchange the end facets of the glass were polished perpendicular to the waveguide resulting in formation of waveguides length ~35 mm. Planar waveguides were characterised using Metricon prism coupler and found to be single mode operating at 1.5 μm and having three modes at 632.8 nm. Effective index at 632.8 nm using transverse electric (TE) polarization of the fundamental mode (zero order mode) is 1.5191, 1.5170 for the first order mode and 1.5157 for the second order mode. At 1.5 μm effective index of the fundamental mode was measured as 1.5093. In addition, channel ion-exchanged waveguides were characterised by imaging the mode profile at the output of the waveguide on IR camera through the x10 focusing length and found to be single mode at 1.5 μm and having three modes at 632.8 nm. Note, since the waveguides are optimised for telecommunication wavelengths, they exhibit surface scattering loss for visible light. This is why they can be visualised with green and red lasers as photographed in Figure 2b. The near field mode profiles were measured to estimate dimensions of the mode profile from our device compared to the mode of SM1550 fibre at the same distance from IEEE-1394 Digital Camera using x10 objective and neutral density (ND20) filter to protect the camera from saturation. Mode profile of SM1550 fibre is about 9.3 μm. The width and depth of the mode profile of our device were found as 5.2565 μm and 7.6826 μm respectively.

**Solvent cleaning**.
Waveguides have been cleaned following standard solvent cleaning routine while increasing solvent polarity. Waveguides have been sonicated for 20 min ultrasonic bath, and then rinsed with isopropanol then with deionised water and blow dried with air gun.

**Surface modification**.
Waveguides were cleaned from organic residues following solvent cleaning routine (see above) and negatively charged using Tepla 300 Plasma Asher machine at 1000 *W* with gas $O_2$ of 600 *ml*/min until the temperature of 145°C was reached. The process took about 5-7 min. The power then was turned off to cool down the waveguides to 55°C. The procedure has been repeated once more in order to generate higher oxygen anions charge on the surface of the waveguides. This method yields a very hydrophilic surface.

**Fabrication of the liquid reservoir**.
To conduct the measurements of liquid solutions on waveguides, the polydimethylsiloxane (PDMS) liquid reservoir has been fabricated. The Sylgard 184 elastomer and curing agent were mixed together at a weight ratio of 10:1. Liquid PDMS has been placed in vacuum conditions for 30 min, and then baked at 90°C for 1 hr, resulting in a robust chamber. Bath-like shape has been formed using aluminium foil supported by a microscope glass. By inspecting the spectrum of the PDMS we confirmed that it is a non-absorbing material at the wavelengths of interest. A demountable PDMS chamber of length 8 mm was self-stick on top of the waveguides to contain the liquid. After 45 min of measurements, PDMS chamber was detached from the waveguide and the droplet of ~3 mm in diameter remained on the surface.

**Optical spectroscopy on a waveguide**.
Optical waveguide measurements of transmittance spectra were conducted using the apparatus shown in Figure 2a. The measurements were realised using the infra-red light generated by a high power fibre continuum source (Fianium SC-600-FC) which is operating at the central wavelength of 1060 nm with spectral bandwidth spanning from 450 nm to beyond 1750 nm and generating optical pulses with duration less than 10 ps. Broadband light was collimated and focused onto polarisation maintaining fibre (PMF) using Melles Grot objective NA 0.65 and magnification x40 to control polarization in the system. PMF was directly fibre-coupled into a channel waveguide (with coupling efficiency of 8 dB/nm) and the power transmitted through the waveguide was fibre-coupled using graded index multimode fibre into an optical spectrum analyser (OSA, Yokogawa AQ6370). Spectral resolution was set to 0.5 nm. The acquisition time for the 300 nm spectral window is about 50 sec. No polarization dependency was confirmed. NMA (index of $n_{NMA}$=1.57118) was diluted in Hex (index of $n_{Hex}$=1.37508) ratio 2:3 resulting in the mixture index of 1.46769. Note: indices of the liquids were measured using RA 510 refractometer operating at 589 nm at room temperature of 21±2°C. At the beginning of the measurement, a droplet of NMA in Hex volume 0.1 *mL* has been dripped onto a PDMS liquid reservoir placed on a waveguide for the chemical analysis. Demonstration of the dynamics of NMA on a guide lasted for 42 min. To conduct the measurements with Aniline, prior the measurement, the waveguides were cleaned and charged by oxygen anions in Tepla 300 Plasma Asher machine.

Electronic address: *A.Karabchevsky@soton.ac.uk; ‡A.Kavokin@soton.ac.uk

**Acknowledgments**
A.V.K. acknowledges support from the EPSRC established career fellowship and A.K. acknowledges support from the BGU (Israel) an Outstanding Woman in Science Award. We thank J. J. Greffet, E. Plum and N. I. Zheludev for helpful discussions.

**Author contributions**
A.K. proposed the idea of the paper, performed sample fabrication, characterised the samples, built experimental setup, performed measurements and theoretical calculations. A.V.K. proposed the interpretation of the results and developed the theoretical aspects. A.K. and A.V.K. analysed the results and co-wrote the paper.

**Competing financial interests**
The authors declare no competing financial interests.